\documentclass[journal,draftcls,onecolumn,12pt,twoside]{IEEEtranTCOM}

\makeatletter
\let\NAT@parse\undefined
\makeatother

\usepackage[numbers,sort&compress]{natbib}
   
\usepackage{amssymb}
\usepackage{graphicx}
\usepackage[cmex10]{amsmath}
\usepackage{mathtools}
\usepackage{color}
\usepackage{float}
\usepackage[nolist,nohyperlinks]{acronym}

\begin{acronym}
\acro{CGQFs}{Complex Gaussian quadratic forms}
\acro{PDF}{probability density function}
\acro{CDF}{cumulative density function}
\acro{MRC}{maximal ratio combining}
\acro{MGF}{moment generating function}
\acro{MSE}{mean squared error}
\acro{BER}{bit error rate}
\acro{SNR}{signal-to-noise ratio}
\acro{LoS}{line of sight}
\acro{MIMO-MRC}{multiple-input multiple-output maximum-ratio combining}
\end{acronym}

\newtheorem{proposition}{Proposition}
\newtheorem{lemma}{Lemma}
\newtheorem{definition}{Definition}

\makeatletter
\def\blfootnote{\xdef\@thefnmark{}\@footnotetext}
\makeatother

\makeatletter
\newcases{mycases}{\quad}{%
  $\m@th\displaystyle{##}$}{$\m@th\displaystyle{##}$\hfil}{\lbrace}{.}
\newcases{eqsyst}{}{%
  $\m@th\displaystyle{##}$}{$\m@th\displaystyle{##}$\hfil}{\lbrace}{.}
\makeatother

\begin{document}

\title{A New Approach to the Statistical Analysis of Non-Central Complex Gaussian Quadratic Forms with Applications}

\author{Pablo Ram\'irez-Espinosa, Laureano Moreno-Pozas, Jos\'e F. Paris, Jos\'e A. Cort\'es and Eduardo Martos-Naya}

\maketitle

\blfootnote{\noindent P. Ram\'irez-Espinosa, J. F. Paris, J.A. Cort\'es and E. Martos-Naya are with Departmento de Ingenier\'ia de Comunicaciones, Universidad de Malaga - Campus de Excelencia Internacional Andaluc\'ia Tech., Malaga 29071, Spain. E-mail: \{pre, paris, jaca, eduardo\}@ic.uma.es. 
 \\ \indent L. Moreno-Pozas is with Department of Electronic and Computer Engineering, School of Electrical and Computer Engineering, Hong Kong University of Science
and Technology, Kowloon, Hong Kong. E-mail: eelaureano@ust.hk.
 \\ \indent
 This work has been funded by the Spanish Government
and the European Fund for Regional Development FEDER (project TEC2014-57901-R).
\\ \indent 
This work has been submitted to the IEEE for publication. Copyright may be transferred without notice, after which this version may no longer be accesible.

}

\begin{abstract}
This paper proposes a novel approach to the statistical characterization of non-central complex Gaussian quadratic forms (CGQFs). Its key strategy is the generation of an auxiliary random variable (RV) that replaces the original CGQF and converges in distribution to it. The technique is valid for both definite and indefinite CGQFs and yields simple expressions of the probability density function (PDF) and the cumulative distribution function (CDF) that only involve elementary functions. This overcomes a major limitation of previous approaches, where the complexity of the resulting PDF and CDF does not allow for further analytical derivations. Additionally, the mean square error between the original CGQF and the auxiliary one is provided in a simple closed-form formulation. These new results are then leveraged to analyze the outage probability and the average bit error rate of maximal ratio combining systems over correlated Rician channels.   
\end{abstract}


\section{Introduction}
\ac{CGQFs} play an essential role when analyzing several wireless techniques, including \ac{MRC} \cite{Rao01}, optimum combining \cite{Lao03}, beamforming \cite{Kim13}, multibeam strategies \cite{Schlegel1996}, orthogonal space time block coding \cite{Ropokis08}, relays \cite{Havary10}, non-coherent modulations \cite{Raphaeli96}, diferential detections \cite{Pauli2008} and matched-field processing \cite{Gall2014}.

The analysis of CGQFs has been usually restricted to central CGQFs, i.e. quadratic forms built from zero mean complex Gaussian vectors, which can be given in a very tractable form \cite{Moustakas16, Provost92}. However, the analysis of non-central \ac{CGQFs} remains as an open problem in the literature. Hence, despite its interest in common problems like the study of digital communications over Rician channels, no closed-form expressions are known for chief probability functions like the \ac{PDF} and the \ac{CDF}, for which only approximated solutions have been given.  

The statistical analysis of non-central \ac{CGQFs} can be traced back to the work by Turin \cite{Turin60}. Although their characteristic function  was given in closed-form, Turin highlighted the challenge of obtaining the \ac{PDF} of \ac{CGQFs} built from non-zero mean Gaussian vectors. Since then, some works made initial progress to pave the way for the complete statistical characterization of non-central \ac{CGQFs}. To the best of the authors' knowledge, most of the approaches available in the literature are based on the direct inversion of the \ac{MGF}, or equivalently, the  characteristic function, to obtain an approximation of the \ac{PDF} of \ac{CGQFs} \cite{Tziritas87, Raphaeli96Taylor, Biyari93}. Some works apply different series expansions to the characteristic function to allow such inversion \cite{Tziritas87, Raphaeli96Taylor}, while the work by Biyari and Lindsey considers a specific non-central CGQF and inverts its \ac{MGF} by solving some convolution integrals \cite{Biyari93}. All these works present approximations for the \ac{PDF} of non-central \ac{CGQFs} in terms of double infinite sum of special functions. In particular, the \ac{PDF} of positive-definite non-central \ac{CGQFs} is given in terms of a double infinite sum of modified Bessel functions in \cite{Tziritas87}, while the \ac{PDF} of indefinite non-central \ac{CGQFs} is expressed in terms of a double infinite sum of incomplete gamma functions \cite{Raphaeli96Taylor} and of a double infinite sum of Laguerre polynomials \cite{Biyari93}. 

Taking into account the limitations of direct inversion methods, since the solutions provided for the PDF and the CDF of non-central CGQFs are difficult to compute and not suitable for any further insightful analysis, very recently, Al-Naffouri et al. presented a different approach. They applied a transformation to the inequality that defines the \ac{CDF} of non-central \ac{CGQFs}, yielding a problem in which the well-known saddle point technique allows expressing the \ac{CDF} as the solution of a differential equation \cite{Moustakas16}.

This paper proposes a completely different approach to the statistical analysis of indefinite non-central \ac{CGQFs}, which leads to simple expressions that approximate both the \ac{PDF} and \ac{CDF} of \ac{CGQFs}. It is based on appropriately perturbing the non-zero mean components of the Gaussian vectors that build the quadratic form. This yields an auxiliary CGQF, denoted as confluent CGQF, which converges in distribution to the original quadratic form and whose analysis is surprisingly simpler. Specifically, this novel approach offers the following advantages over the recently proposed work in \cite{Moustakas16} and the other approaches given in the literature \cite{Tziritas87, Raphaeli96Taylor, Biyari93}:

\begin{itemize}
	\item The probability functions, namely \ac{PDF} and \ac{CDF}, are given as a linear combination of elementary functions (exponentials and powers) in a very tractable form. 
	\item Simple closed-form expression for the \ac{MSE} between the CGQF and the auxiliary one is provided, allowing the particularization of the auxiliary variable in order to make this error drop below a certain threshold. 
	\item Since the statistics of an auxiliary random variable are used to characterize \ac{CGQFs}, the approximated solution is also a statistical distribution. This is not the case of the infinite series expressions of the literature, which are no longer strict \ac{PDF}s when truncated (they have no-unit area).  
\end{itemize}

Finally, with the aim of exemplifying the tractability of the derived expressions, they are used to further study the performance of \ac{MRC} systems over non-identically distributed Rician fading channels with arbitrary correlation. Hence, simple expressions for the outage probability and the \ac{BER} are provided for different modulation schemes. 

The remainder of this paper is structured as follows. The notation and some preliminary results are introduced in Section \ref{sec:Notation}. Section \ref{sec:StatChar} presents the general approach, as well as the statistical characterization of indefinite non-central \ac{CGQFs} with a very simple and precise approximation which admits a closed-form expression for its associated \ac{MSE}. In Section \ref{sec:Residues}, an efficient recursive algorithm to compute the derived expressions is introduced. In Section \ref{sec:MRC}, the new statistical characterization of non-central \ac{CGQFs} is applied to the performance analysis of \ac{MRC} systems over correlated Rician channels. Finally, conclusions are drawn in Section \ref{sec:Conclusion}.

\section{Notation and Background}
\label{sec:Notation}

Throughout this paper, the following notation will be used. Vectors and matrices are denoted in
bold lowercase and bold uppercase, respectively. $\mathbb{E}[\cdot]$ is the expectation operator, while $\mathcal{L}\{\cdot\}$ and $\mathcal{L}^{-1}\{\cdot\}$ denote the Laplace transform and the inverse Laplace transform operators, respectively. The symbol $\sim$ signifies \emph{statistically distributed as}. The superscript $(\cdot)^\dagger$ indicates matrix complex conjugate transpose and ${\rm tr}(\cdot)$ is the matrix trace. The matrices $\mathbf{I}_p$ and $\mathbf{0}_{p\times q}$ denote a $p\times p$ identity and a $p\times q$ all-zero matrix, respectively. When ${\rm diag}(\cdot)$ is applied to a matrix, it returns a vector whose entries are the diagonal elements of that matrix.  Additionally, ${\rm u}(\cdot)$ is the unit step function whose value is $1$ if the argument is non-negative and $0$ otherwise. ${\rm sgn}(\cdot)$ is the sign function whose value is $1$ for non-negative arguments and $-1$ otherwise. Some relevant definitions and preliminary results, which will be used when presenting the main contributions, are now introduced.

\subsection{Basic distributions}

\begin{definition}[Gamma distribution]
	Let $X$ be a real random variable, which follows a Gamma distribution with shape parameter $m$ and scale parameter $\theta$, i.e. $X\sim\Gamma\left(m,\theta\right)$. Then, the PDF of $X$ is given by
	\begin{equation}
		\label{eq:GammaPDF}
		f_X(x) = \frac{1}{\Gamma(m)\theta^m}x^{m-1}e^{-x/\theta}
	\end{equation}
	where $\Gamma(\cdot)$ is the gamma function and $m, \theta\in\mathbb{R}^+$.
\end{definition}

\begin{definition}[Non-central $\chi^2$ distribution with $n$ degrees of freedom]

Define
\begin{equation}
		Y = \sum_{i=1}^n X_i^2
	\end{equation}
where $X_i$, $i = 1,\dots , n$, are statistically independent real Gaussian random variables with unit variance and means $\mu_i$, i.e. $X_i\sim\mathcal{N}\left(\mu_i,1\right)$. Then, $Y$ follows a non-central $\chi^2$ distribution with $n$ degrees of freedom and noncentrality parameter $\delta = \sum\limits_{i=1}^n\mu_i^2$, i.e. $Y\sim\chi_n^2\left(\delta\right)$. The MGF of $Y$ is therefore given by
	\begin{equation}
		\label{eq:MGFchi}
		M_Y(s) = \frac{1}{\left(1-2s\right)^{n/2}}\exp\left(\frac{\delta s}{1-2s}\right).
	\end{equation}
\end{definition}

\subsection{Complex Gaussian Quadratic Forms} 
\begin{definition}[CGQF]
	Let $\mathbf{v}\in \mathbb{C}^{n\times 1}$ be a random vector that follows a $n$-variate Gaussian distribution with mean vector $\mathbf{\overline{v}}\in\mathbb{C}^{n\times 1}$ and non-singular Hermitian covariance matrix $\mathbf{L}\in \mathbb{C}^{n\times n}$, i.e. $\mathbf{v}\sim \mathcal{CN}_n\left(\mathbf{\overline{v}},\mathbf{L}\right)$, and let $\mathbf{A}\in\mathbb{C}^{n\times n}$ be a non-singular indefinite Hermitian matrix, i.e. $\mathbf{A}$ can have positive and negative real eigenvalues. Then, the real random variable
	\begin{equation}
		\label{eq:QF}
		Q = \mathbf{v}^{\dagger}\mathbf{Av}
	\end{equation}
is an indefinite non-central CGQF.
\end{definition}

Note that assuming $\mathbf{L}$ is non-singular does not implies any loss of generality. If ${\rm rank}(\mathbf{L})<n$, then some components of $\mathbf{v}$ are linearly related \cite[chap. 1]{Muirhead2009} and, consequently, $Q$ can be rewritten in terms of another vector with full-rank covariance matrix. That is equivalent to assuming $\mathbf{L}$ is positive definite.

The expression in (\ref{eq:QF}) has been classically employed in the context of CGQF analysis \cite{Turin60, Slichenko2014, Raphaeli96Taylor}. However, in this work, an alternative form will be used, which is equivalent to (\ref{eq:QF}) and can be deduced from it by performing some algebraic transformations. 
This alternative form is formally equivalent to the one in \cite[eq. (2)]{Tziritas87}, but it will be derived using a different approach that facilitates the understanding of the subsequent analysis proposed in this paper.

Since the covariance matrix $\mathbf{L}$ is a positive definite Hermitian matrix, a Cholesky factorization is performed such that $\mathbf{L} = \mathbf{CC}^{\dagger}$, where $\mathbf{C} \in \mathbb{C}^{n\times n}$ is an invertible lower triangular matrix with non-negative diagonal entries \cite{Horn90}. Then, the vector $\mathbf{v}$ can be expressed as
\begin{equation}
	\label{eq:vChol}
	\mathbf{v} = \mathbf{Cz} + \mathbf{\overline{v}}
\end{equation}
where $\mathbf{z}\sim\mathcal{CN}_n\left(\mathbf{0}_{n\times 1},\mathbf{I}_n\right)$. Substituting \eqref{eq:vChol} in \eqref{eq:QF} and after some algebraic manipulations, one has
\begin{equation}
	\label{eq:Qint2}
	Q = \left(\mathbf{z} + \mathbf{C}^{-1}\mathbf{\overline{v}}\right)^{\dagger}\mathbf{C}^{\dagger}\mathbf{AC}\left(\mathbf{z} + \mathbf{C}^{-1}\mathbf{\overline{v}}\right).
\end{equation} 

$\mathbf{C}^{\dagger}\mathbf{AC}$ is Hermitian, so it can be diagonalized as
\begin{equation}
	\mathbf{C}^{\dagger}\mathbf{AC} = \mathbf{U\Lambda U}^{\dagger}
\end{equation}
where $\mathbf{U}$ is an unitary matrix and $\mathbf{\Lambda}$ is a diagonal matrix whose entries, $\lambda_i$ for $i=1,\dots,n$, are the eigenvalues of $\mathbf{C}^{\dagger}\mathbf{AC}$ (or, equivalently, those of $\mathbf{LA}$) \cite[chap. 2]{Scharf91}. Thus, relabeling $\mathbf{y} = \mathbf{U}^{\dagger}\mathbf{z}$ and $\mathbf{\overline{h}} = \mathbf{U}^{\dagger}\mathbf{C}^{-1}\mathbf{\overline{v}}$, one gets
\begin{equation}
	\label{eq:QFDiag}
	Q = \left(\mathbf{y}+\mathbf{\overline{h}}\right)^{\dagger}\mathbf{\Lambda}\left(\mathbf{y}+\mathbf{\overline{h}}\right).
\end{equation}

Since $\mathbf{U}$ is unitary, the distribution of $\mathbf{y}$ is the same as that of $\mathbf{z}$, i.e. $\mathbf{y}\sim\mathcal{CN}_n\left(\mathbf{0}_{n\times 1},\mathbf{I}_n\right)$. Consequently, the quadratic form $Q$ is now expressed in terms of a random vector $\mathbf{y}$ whose elements are independent and the diagonal matrix $\mathbf{\Lambda}$ with the eigenvalues of $\mathbf{LA}$. Depending on whether $Q$ is definite or indefinite, all the eigenvalues have the same sign or not. For positive definite and negative definite CGQFs, $\lambda_i > 0$ and $\lambda_i < 0$ for $i=1,\dots,n$, respectively. In turn, when $Q$ is indefinite, the eigenvalues can be either positive or negative. In order to obtain the MGF of $Q$, \eqref{eq:QFDiag} is expanded as
\begin{equation}
	Q = \sum_{i=1}^{n}\lambda_i\left(y_i+\overline{h}_i\right)^{\dagger}\left(y_i+\overline{h}_i\right)
\end{equation}  
where $y_i$ and $\overline{h}_i$, with $i=1,\dots,n$, are the entries of $\mathbf{y}$ and $\mathbf{\overline{h}}$, respectively. Additionally, defining $Y_i = 2\left(y_i+\overline{h}_i\right)^{\dagger}\left(y_i+\overline{h}_i\right)$, the statistical independence of the elements of $\mathbf{y}$ allows expressing $Q$ as
\begin{equation}
	\label{eq:QFChi}
	Q = \sum_{i=1}^{n}\frac{\lambda_i}{2}Y_i
\end{equation}
with $Y_i\sim\chi_2^2\left(2\left|\overline{h}_i\right|^2\right)$. Thus, $Q$ can be expressed in terms of a linear combination of independent non-central $\chi^2$ variables. Hence, its MGF can be straightforwardly obtained as the product of the MGFs of the scaled version of $Y_i$, which can be deduced from \eqref{eq:MGFchi}, getting the result given in \cite[eq.~(7)]{Tziritas87} 
\begin{equation}
		\label{eq:MGFCGQF}
		M_Q(s) = \prod_{i=1}^n\frac{\exp\left(\frac{\lambda_i \mu_i s}{1-\lambda_i s}\right)}{1-\lambda_i s}
	\end{equation}	 
where $\mu_i = \left|\overline{h}_i\right|^2 = \left[\mathbf{U}^{\dagger}\mathbf{C}^{-1}\mathbf{\overline{v}\overline{v}}^{\dagger}\left(\mathbf{C}^{-1}\right)^{\dagger}\mathbf{U}\right]_{i,i}$. Note that the distinct sign with respect to \cite[eq.~(7)]{Tziritas87} is due to a slightly different definition of the MGF, which in \eqref{eq:MGFCGQF} is calculated as $M_Q(s) = \mathbb{E}\left[e^{sQ}\right]$.

Closed-form expressions for some statistics of $Q$, e.g., the PDF and CDF, are not known due to the exponential term in \eqref{eq:MGFCGQF}, which considerably complicates performing an inverse Laplace transformation.  
 This issue is a direct consequence of considering a non-central Gaussian \mbox{vector $\mathbf{v}$}. Actually, this does not occur when $\mathbf{v}$ has zero mean, since the exponential term in \eqref{eq:MGFCGQF} vanishes, allowing a straightforward inversion to obtain the distribution of $Q$. Although the idea behind most previous contributions consists in expanding such exponential function to perform the inverse Laplace transform, the approach here presented will circumvent the need of manipulating this function, which usually leads to complicated statistical expressions that are not suitable for subsequent analyses \cite{Tziritas87, Raphaeli96Taylor, Biyari93}. It is based on randomly perturbing the deterministic elements that originate the non-centrality of the quadratic form, such that the exponential term in \eqref{eq:MGFCGQF} disappears, thus facilitating the derivation of the distribution of $Q$. This approach will be referred to as principle of confluence in the next section.
\section{Confluent Non-Central Complex Gaussian Quadratic Form}
\label{sec:StatChar}

The here proposed approach exploits the fact that the analysis of some statistical problems is notably simplified by introducing a random fluctuation into them. A major innovation of this paper is the determination of an adequate fluctuation to achieve this end in the context of non-central \ac{CGQFs}. To this aim, an auxiliary CGQF is obtained by perturbing vector $\mathbf{\overline{h}}$ in \eqref{eq:QFDiag} with a random variable that depends on a shape parameter. When the latter tends to infinity, the auxiliary CGQF converges to the original one. This section firstly formalizes this property, referred to as principle of confluence, and then shows that the statistical analysis of the original CGQF can be derived from that of the auxiliary one. 

\subsection{The Principle of Confluence}
\label{sec:ConfPrinc}

In the following, the definition of confluent random variable is given, as well as some relevant lemmas that build a general framework that will be used to analyze non-central \ac{CGQFs}.

\begin{definition}[Confluent random variable]
	\label{def:Confluence}
	Let $X_m$ be a real random variable with a real and positive shape parameter $m$. Then, the random variable $X_m$ is \textit{confluent} in $m$ to another random variable $X$ if
	\begin{equation}
		\lim_{m\to\infty}M_{X_m}(s) = M_X(s)
	\end{equation}
for $s = it$ with $i=\sqrt{-1}$ and $t\in\mathbb{R}$, and it is denoted as $X_m \leadsto X$. Also, $X$ is named the \textit{limit} variable.
\end{definition}

\begin{definition}[Weak convergence]
	Let $\{X_n\}$ be a sequence of real random variables. Then, $\{X_n\}$ is said to converge weakly, or converge in distribution, to another random variable $X$ if
	\begin{equation}
		\lim_{n\to\infty}F_{X_n}(x) = F_X(x)
	\end{equation}	 
at every continuity point \cite{Billingsley2012}, where $F_X(x)$ is the \ac{CDF} of $X$.
\end{definition}

With the above definitions, the following lemmas are now presented.

\begin{lemma}
	\label{lem:WeakConv}
	Let $X_m$ and $X$ be two random variables such that $X_m \leadsto X$. Then, $X_m$ converges weakly to $X$ in $m$.
\end{lemma}
\begin{IEEEproof}
	The lemma is a consequence of L\'evy's continuity theorem (or L\'evy's convergence theorem) \cite[chap. 18]{Williams91}. The confluence in $m$ between the random variables $X_m$ and $X$ implies that the MGF of $X_m$ converges pointwise to $M_X(s)$ in the imaginary axis. This is equivalent to the convergence of the characteristic functions, fulfilling L\'evy's theorem and ensuring the weak convergence of $X_m$ to $X$.
\end{IEEEproof}

\begin{lemma}
	\label{lem:ExpConv}
	Let $X_m$ and $X$ be two random variables such that $X_m \leadsto X$ in $m$. If $g: \mathbb{R}\to \mathbb{R}$ is a continuous and bounded function, then
	\begin{equation}
		\lim_{m\to\infty}\mathbb{E}\left[g(X_m)\right] = \mathbb{E}\left[g(X)\right].
	\end{equation}
\end{lemma}
\begin{IEEEproof}
	As $X_m\leadsto X$, Lemma \ref{lem:WeakConv} ensures the weak convergence between $X_m$ and $X$. According to Helly-Bray theorem, this weak convergence is equivalent to the convergence in expectations if and only if $g$ is a continuous and bounded function \cite{Manoukian1986}.
\end{IEEEproof}

Definition \ref{def:Confluence} along with Lemma \ref{lem:WeakConv} and Lemma \ref{lem:ExpConv} constitute the principle of confluence. It allows circumventing the need of manipulating the statistics of X, working with those of $X_m$ instead. 
As such, the principle of confluence is a novel approach to characterize complicated random variables, by means of auxiliary variables which are more tractable to analyze.

\subsection{The Principle of Confluence for CGQFs}
\label{subsec:ConfCGQF}

The principle of confluence is here used to analyze non-central \ac{CGQFs}. In order to do so, it is necessary to firstly define the auxiliary random variable, which will be referred to as confluent CGQF.
\begin{proposition}
	Let $\xi_{m,i}$ for $i=1,\dots,n$ be a set of non-negative random variables such that $\xi_{m,i}^2\sim\Gamma(m,1/m)\;\forall\;i$. Then, $\xi^2_{m,i}\leadsto 1$ and $\xi_{m,i}\leadsto 1$ in $m\;\forall\;i$. 
\end{proposition}
\begin{IEEEproof}
The confluence of $\xi_{m,i}^2$ is straightforwardly proved since $\lim\limits_{m\to\infty}M_{\xi_{m,i}^2}(s)= e^s$. The confluence of $\xi_{m,i}$ can be proved from the relationship between its \ac{CDF} and that of $\xi^2_{m,i}$.
\end{IEEEproof}
\begin{proposition}
	\label{pro:MGF}
	Consider $Q$ the non-central CGQF in \eqref{eq:QFDiag} and let $\mathbf{D_{\xi}}\in \mathbb{R}^{n\times n}$ be a diagonal matrix with entries $\xi_{m,i}$ for $i=1,\ldots,n$. Then, 
	\begin{equation}
		\label{eq:QFconf}
		Q_m = \left(\mathbf{y}+\mathbf{D_{\xi}}\mathbf{\bar{h}}\right)^{\dagger}\mathbf{\Lambda}\left(\mathbf{y}+\mathbf{D_{\xi}}\mathbf{\bar{h}}\right)
	\end{equation}
is a confluent non-central CGQF whose \ac{MGF} is given by 
	\begin{equation}
		\label{eq:MGF}
		M_{Q_m}(s) = \prod_{i=1}^n\frac{\left(1-\frac{\lambda_i\mu_i s}{m\left(1-\lambda_i s\right)}\right)^{-m}}{1-\lambda_i s}
	\end{equation}
	where $\lim\limits_{m\to\infty}M_{Q_m}(s) = M_Q(s)$ and, therefore, $Q_m\leadsto Q$. 
\end{proposition}
\begin{IEEEproof}
	The \ac{MGF} in (\ref{eq:MGF}) is derived in Appendix \ref{app:MGFproof} and, since 
\begin{equation}
	\label{eq:Limit}
	\lim_{m\to\infty}\left(1-\frac{\lambda_i\mu_i s}{m\left(1-\lambda_i s\right)}\right)^{-m} = \exp\left(\frac{\lambda_i \mu_i s}{1-\lambda_i s}\right),
\end{equation}
then $\lim\limits_{m\to\infty}M_{Q_m}(s) = M_Q(s)$, which proves that $Q_m$ is confluent in $m$ to $Q$.
\end{IEEEproof}

To give an intuitive explanation for Proposition \ref{pro:MGF}, observe \eqref{eq:QFconf}. Since $\xi_{m,i} \leadsto 1$ for $i=1,\dots,n$, $\mathbf{D_{\xi}}$ becomes the identity matrix when $m\to\infty$. Consequently, the confluent CQGF in \eqref{eq:QFconf} becomes the original one in \eqref{eq:QFDiag} in the limit. Additionally, the expression in (\ref{eq:MGF}) confirms that choosing the perturbing fluctuations to follow Gamma distributions is appropriate since this expression does no longer present the exponential term. In fact, by performing some algebraic manipulations in \eqref{eq:MGF}, the \ac{MGF} of $Q_m$ can be written as 
\begin{equation}
	\label{eq:MGFRs}
	M_{Q_m}(s) = \prod_{k=1}^n\left[(-\lambda_k)\left(1+\frac{\mu_k}{m}\right)^m\right]^{-1}\prod_{i=1}^n\frac{\left(s-1/\lambda_i\right)^{m-1}}{\left(s-\beta_i\right)^m}
\end{equation}
where $M_{Q_m}(s)$ is in terms of a rational polynomial whose zeroes and poles are $1/\lambda_i$ and \mbox{$\beta_i = \left[\lambda_i\left(1+\mu_i/m\right)\right]^{-1}$} for $i=1,\dots,n$, respectively. Assuming there can be repeated poles and zeroes, and taking into account that, if $\mu_i=0$ for a certain $i$, then $\beta_i = 1/\lambda_i$, this rational polynomial can be simplified. Thus, denoting as $\widetilde{\beta}_i$ and $1/\widetilde{\lambda}_j$ for $i=1,\dots,n_\beta$ and $j = 1,\dots,n_\lambda$ the distinct poles and zeroes resulting from simplifying the rational polynomial in \eqref{eq:MGFRs} with multiplicities $p_i$ and $q_i$, respectively, the \ac{MGF} of $Q_m$ is expressed
\begin{equation}
	\label{eq:MGFRsSimplf}
	M_{Q_m}(s) = \prod_{k=1}^n\left[(-\lambda_k)\left(1+\frac{\mu_k}{m}\right)^m\right]^{-1}\frac{\prod\limits_{j=1}^{n_\lambda}\left(s-1/\widetilde{\lambda}_j\right)^{q_j}}{\prod\limits_{i=1}^{n_{\beta}}\left(s-\widetilde{\beta}_i\right)^{p_i}}.
\end{equation}

In contrast with the \ac{MGF} of the original CGQF in \eqref{eq:MGFCGQF}, the expression of $M_{Q_m}(s)$ in \eqref{eq:MGFRsSimplf} allows a straightforward inversion, i.e. performing an inverse Laplace transformation in order to obtain the \ac{PDF} and the \ac{CDF} of $Q_m$. Moreover, from Proposition \ref{pro:MGF}, since $Q_m\leadsto Q$, the statistics of $Q$ can be obtained from those of $Q_m$ by virtue of Lemma \ref{lem:WeakConv}. Thus, the \ac{PDF} and \ac{CDF} of $Q_m$ are calculated in the following propositions, which are easily derived from \eqref{eq:MGFRsSimplf} after expanding the rational polynomial in partial fractions.
\begin{proposition}
	\label{pro:PDF}
	Consider $Q_m$ the confluent CGQF  defined in \eqref{eq:QFconf}. Then, the \ac{PDF} of $Q_m$ is given by  a linear combination of elementary functions as
	\begin{equation}
		\label{eq:PDF}
		f_{Q_m}(x) =\sum_{i=1}^{n_{\beta}}\sum_{j=1}^{p_i}\alpha_{i,j}e^{-\widetilde{\beta}_i x}x^{j-1}{\rm u}\left(\widetilde{\beta}_i x\right){\rm sgn}(x)
	\end{equation}	
	where $\alpha_{i,j} = B_jA_{i,j}$ with
\begin{equation}
	B_j = \frac{1}{(j-1)!}\prod_{k=1}^n\left[(-\lambda_k)\left(1+\frac{\mu_k}{m}\right)^m\right]^{-1}
\end{equation}
	and $A_{i,j}$ are the residues that arises from performing a partial fraction decomposition in \eqref{eq:MGFRsSimplf} after evaluating $M_{Q_m}(-s)$. A closed-form expression for $A_{i,j}$ is given by
\begin{equation}
\label{eq:Aij}
	A_{i,j} = \sum_{\substack{k_1+...+k_{N-1}=p_i-j\\k_u \leq q_u, u=1,\dots,n_\lambda}}\prod\limits_{s=1}^{N-1}(k_s!)^{-1}D_i(k_1,\dots,k_{N-1})
\end{equation}
with 
\begin{equation}
	D_i(k_1,\dots,k_{N-1}) = \prod_{t=1}^{n_\lambda}{\frac{\left(\widetilde{\lambda}_t^{-1}-\widetilde{\beta}_i\right)^{q_t-k_t}}{(q_t!)^{-1}(q_t-k_t)!}} \prod_{r=1}^{i-1}{\frac{(-1)^{k_{r+n_\lambda}}(p_r)_{k_{r+n_\lambda}}}{\left(\widetilde{\beta}_r-\widetilde{\beta}_i\right)^{p_r+k_{r+n_\lambda}}}}\prod_{l=i+1}^{n_{\beta}}{\frac{(-1)^{k_{l+n_\lambda-1}}(p_l)_{k_{l+n_\lambda-1}}}{\left(\widetilde{\beta}_l-\widetilde{\beta}_i\right)^{p_r+k_{l+n_\lambda-1}}}}
\end{equation}
as proved in Appendix \ref{app:Residues}, where the sum in \eqref{eq:Aij} is over all possible combinations of $k_1,...,k_{N-1}$, with $N = n_\lambda + n_\beta$ and $k_u \leq q_u$ for $u=1,\dots,n_\lambda$, that meet \mbox{$\sum\limits_{t=1}^{N-1}k_t=p_i-j$}.
\end{proposition}
\begin{IEEEproof}
	The \ac{PDF} of $Q_m$ is easily obtained from \eqref{eq:MGFRsSimplf} as $f_{Q_m}(x) = \mathcal{L}^{-1}\left\{M_{Q_m}(-s)\right\}$ by performing a partial fraction expansion as detailed in Appendix \ref{app:PDFproof}.
\end{IEEEproof}

\begin{proposition}
	\label{pro:CDF}
	Consider $Q_m$ the confluent CGQF  defined in \eqref{eq:QFconf}. Then, the \ac{CDF} of $Q_m$ is given by 
	\begin{equation}
		\label{eq:CDF}
		F_{Q_m}(x) = {\rm u}(x) + \sum_{i=1}^{n_{\beta}}\sum_{j=1}^{p_i}\omega_{i,j}e^{-\widetilde{\beta}_i x}x^{j-1}{\rm u}\left(\widetilde{\beta}_ix\right){\rm sgn}(x)
	\end{equation}
where $\omega_{i,j} = B_j C_{i,j}$ with $C_{i,j}$ the partial expansion residues given by
\begin{equation}
\label{eq:Cij}
	C_{i,j} = \sum_{\substack{k_1+...+k_N=p_i-j\\k_u\leq q_u,u=1,\dots,n_\lambda}}\prod\limits_{s=1}^N(k_s!)^{-1} \frac{k_N!(-1)^{k_N}}{\left(-\widetilde{\beta}_i\right)^{1+k_N}} D_i(k_1,\dots,k_{N-1})
\end{equation}
as can be deduced from Appendix \ref{app:Residues}.
\end{proposition}
\begin{IEEEproof}
	Following the same steps as in the previous proof, the CDF of $Q_m$ is straightforwardly calculated from \eqref{eq:MGFRsSimplf} as $F_{Q_m}(x) = \mathcal{L}^{-1}\left\{M_{Q_m}(-s)/s\right\}$, as detailed in Appendix \ref{app:CDFproof}.
\end{IEEEproof}

Propositions \ref{pro:PDF} and \ref{pro:CDF} provide simple closed-form expressions for both the \ac{PDF} and \ac{CDF} of $Q_m$ in terms of elementary functions, i.e. exponentials and powers. Regarding the argument of the unit step function, it is clear that the domain of ${\rm u}(\widetilde{\beta_i}x)$ where the function values are non-negative will depend on the sign of $\widetilde{\beta}_i$. Thus, if $x<0$, then the value of ${\rm u}(\widetilde{\beta_i}x)$ will be zero for those $\widetilde{\beta}_i > 0$ for $i = 1,\dots,n_\beta$. In turn, for positive values of $x$, the value of the step function will be zero for those $\widetilde{\beta}_i < 0$. Moreover, since $\mu_i \geq 0 \;\forall\; i$, the sign of $\widetilde{\beta}_i$ is the same as that of $\widetilde{\lambda_i}$, so  the shape of the distribution of $Q_m$ (and, equivalently, that of $Q$) depends on the positive eigenvalues $\widetilde{\lambda_i}$ for $x\geq 0$, while it depends on the negative eigenvalues for $x<0$. 

From \eqref{eq:PDF} and \eqref{eq:CDF}, the \ac{PDF} and \ac{CDF} of $Q$ can be approximated by setting $m$ to a sufficiently large value. Note that, although $m$ does not appear explicitly in \eqref{eq:PDF} and \eqref{eq:CDF}, both the poles $\widetilde{\beta_i}$ and their multiplicities $p_i$ depends on $m$, as well as zeroes multiplicities $q_j$. Additionally, in contrast to previous approximations found in the literature, it is possible to quantify the \ac{MSE} between $Q$ and $Q_m$ in closed-form. This result is given in the following proposition.

\begin{proposition}
	Consider $Q$ the CGQF and $Q_m$ the confluent CGQF given in \eqref{eq:QFDiag} and \eqref{eq:QFconf}, respectively. Then, the \ac{MSE} between $Q$ and $Q_m$ is given by
	\begin{align}
		\overline{\epsilon^2} &\triangleq \mathbb{E}\left[\left(Q_m-Q\right)^2\right]= \mathbb{E}\left[\left( \left(\mathbf{y}+\mathbf{D_{\xi}}\mathbf{\bar{h}}\right)^{\dagger}\mathbf{\Lambda}\left(\mathbf{y}+\mathbf{D_{\xi}}\mathbf{\bar{h}}\right) - \left(\mathbf{y}+\mathbf{\bar{h}}\right)^{\dagger}\mathbf{\Lambda}\left(\mathbf{y}+\mathbf{\bar{h}}\right)\right)^2\right] \notag \\
		&= \sum_{i=1}^n\lambda_i^2\mu_i\left[4\left(1-\frac{\Gamma\left(m+1/2\right)}{m^{1/2}\Gamma(m)}\right)+\frac{\mu_i}{m}\right].
		\label{eq:MSE}
	\end{align}
\end{proposition}
\begin{IEEEproof}
	See Appendix \ref{app:Error}.
\end{IEEEproof} 

Observe that, when $\mu_i = \left|h_i\right|^2 = 0$ for $i=1,\dots,n$ then $\overline{\epsilon^2} = 0$ for any value of $m$. This is coherent with the fact that setting $\mu_i=0 \;\forall\;i$ implies having a central CGQF. Additionally, it is easy to prove that the error also goes to zero when $m\rightarrow\infty$. By applying the asymptotic formula for the gamma function given in \cite[eq. 6.1.39]{Abra72}, which allows to write
\begin{equation}
	\label{eq:GammaAsym}
	\Gamma(m+\tau) \approx \sqrt{2\pi}e^{-m}m^{m+\tau-1/2}
\end{equation}
for large $m$, it is clear that $\lim\limits_{m\to\infty}\overline{\epsilon^2} = 0$. This implies that $Q_m$ converges in mean square to $Q$, which is a more general type of convergence between random variables. In fact, convergence in mean square also implies convergence in probability and, consequently, weak convergence \cite{Resnick1998}.

When compared to other approximations, the novel approach here presented renders more tractable expressions for the chief probability functions of indefinite non-central \ac{CGQFs}. The \ac{PDF} and \ac{CDF} of $Q$ can be approximated from those of $Q_m$, which only involves elementary functions in contrast to the more complicated expressions available in the literature \cite{Moustakas16,Tziritas87, Raphaeli96Taylor, Biyari93}. It is only necessary to set $m$ large enough, such that the \ac{MSE} between $Q_m$ and $Q$ drops below a certain threshold.

\section{Discussion on the Computation of Partial Fraction Expansion Residues}
\label{sec:Residues}
 The expressions of $f_{Q_m}(x)$ and $F_{Q_m}(x)$ have been obtained as the inverse Laplace transformation of $M_{Q_m}(-s)$ and $M_{Q_m}(-s)/s$, respectively. These transformations are performed by expanding the rational polynomial in \eqref{eq:MGFRsSimplf} after evaluating $M_{Q_m}(-s)$, that is
\begin{equation}
	\label{eq:Rs}
 R(s) = \frac{\prod\limits_{t = 1}^{n_\lambda}\left(s+1/\widetilde{\lambda}_t\right)^{q_t}}{\prod\limits_{i = 1}^{n_{\beta}}\left(s+\widetilde{\beta}_i\right)^{p_i}},
\end{equation}
so the constants $\alpha_{i,j}$ and $\omega_{i,j}$ in \eqref{eq:PDF} and \eqref{eq:CDF} depend on the partial expansion residues of $R(s)$ and $R(s)/s$, namely $A_{i,j}$ and $C_{i,j}$, respectively.   

Although closed-form expressions for such constants have been provided in \eqref{eq:Aij}  and \eqref{eq:Cij}, the computation of these expressions is impractical for very large $m$. Because the number of terms that needs to be computed for those residues depends on a combinatorial, it grows exponentially with $p_i$ and, consequently, with $m$. As such, for $m$ sufficiently large, the number of combinations becomes computationally
unbearable. This issue is commonly referred to as combinatorial explosion.

Since the partial expansion residues can be defined as derivatives of the rational polynomial \cite[eq. (A.36)]{Oppenheim96}, an alternative approach to avoid the combinatorial explosion may be the calculation of these derivatives by means of Cauchy{'}s differentiation formula, which allows expressing the derivatives as contour integrals over a closed path \cite{Ahlfors66}. Even though these integrals could be numerically computed, they suffer from significant numerical problems as $m$ increases. These are due to the large amplitude oscillatory behavior (with positive and negative values) of the integrands, which can be tens of orders of magnitude larger than the actual value of the integrals, preventing the integral convergence. 

A more suitable approach for the computation of $A_{i,j}$ and $C_{i,j}$ is the algorithm proposed in \cite{Ma2014}, which provides recursive expressions for the partial fraction residues of both proper and improper rational functions. According to \cite[eq. (11a) and (11b)]{Ma2014}, each residue $A_{i,j}$ for $i=1,\dots,n_\beta$ and $j=1,\dots,p_i$ is calculated as a linear combination of the previous ones. The recursion starts from $A_{i,p_i}$, which can be directly computed from the definition in \cite[eq. (A.36)]{Oppenheim96} without taking any derivative. From it, this algorithm computes $A_{i,p_i-1}, A_{i,p_i-2}, \dots, A_{i,1}$ recursively as 
\begin{equation}
	\label{eq:AijRecur}
	A_{i,j} = \begin{mycases}
		\frac{1}{p_i-j}\sum_{k = 1}^{p_i-j}A_{i,j+k}\rho_A(k,-\widetilde{\beta}_i), & \text{if}\;\; 1\leq j \leq p_i-1 \\[1ex]
		\frac{\prod\limits_{t = 1}^{n_{\lambda}}\left(\widetilde{\lambda}_t^{-1}-\widetilde{\beta}_i\right)^{q_t}}{\prod\limits_{\substack{l = 1\\l\neq i}}^{n_{\beta}}\left(\widetilde{\beta}_l-\widetilde{\beta}_i\right)^{p_l}}, & \text{if}\;\; j = p_i
	\end{mycases}
\end{equation}
where $\rho_A(k,s)$ is given by  
\begin{equation}
	\rho_A(k,s) = \sum_{\substack{l = 1 \\ l \neq i}}^{n_{\beta}}\frac{p_l}{\left(-\widetilde{\beta}_l-s\right)^k} - \sum_{t = 1 }^{n_{\lambda}}\frac{q_t}{\left(-\widetilde{\lambda}_t^{-1}-s\right)^k}.
\end{equation}

Analogously, $C_{i,j}$ for $i=1,\dots,n_\beta$ and $j=1,\dots,p_i$  arises as the partial expansion residues of $R(s)/s$, so following the same steps as with $A_{i,j}$ one has
\begin{equation}
	\label{eq:CijRecur}
	C_{i,j} = \begin{mycases}
		\frac{1}{p_i-j}\sum_{k = 1}^{p_i-j}A_{i,j+k}\rho_C(k,-\widetilde{\beta}_i), & \text{if}\;\; 1 \leq j \leq p_i-1 \\[1ex]
		\frac{\prod\limits_{t = 1}^{n_{\lambda}}\left(\widetilde{\lambda}_t^{-1}-\widetilde{\beta}_i\right)^{q_t}}{-\widetilde{\beta}_i\prod\limits_{\substack{l = 1\\l\neq i}}^{n_{\beta}}\left(\widetilde{\beta}_l-\widetilde{\beta}_i\right)^{p_l}}, & \text{if}\;\; j = p_i
	\end{mycases}
\end{equation}
with
\begin{equation}
	\rho_C(k,s) = \sum_{\substack{l = 1 \\ l \neq i}}^{n_{\beta}}\frac{p_l}{\left(-\widetilde{\beta}_l-s\right)^k} + \frac{1}{(-s)^k} - \sum_{t = 1 }^{n_{\lambda}}\frac{q_t}{\left(-\widetilde{\lambda}_t^{-1}-s\right)^k}.
\end{equation} 

 In contrast to \eqref{eq:Aij} and \eqref{eq:Cij}, the computational cost of \eqref{eq:AijRecur} and \eqref{eq:CijRecur} grows linearly with $m$ instead of exponentially, avoiding the combinatorial explosion. Despite that, numerical errors could still be relevant when computing \eqref{eq:AijRecur} and \eqref{eq:CijRecur} due to the limited floating-point precision in calculation software. For very large $m$, the distinct terms in the summation can still differ in considerable orders of magnitude, which could lead to inaccurate results.  However, in contrast to the previous approach that employs Cauchy formula, this computational issue can be solved by working with rational numbers in the software \textsc{Mathematica}. This suite allows the possibility of working with floating-point number with full precision by rationalizing them using the function \textsc{Rationalize}, allowing an error-free computation of $A_{i,j}$ and $C_{i,j}$.

\section{Practical Example: MRC Systems over Correlated Rician Fading Channels}
\label{sec:MRC}

The usefulness of the novel results is now exemplified through the performance analysis of \ac{MRC} over correlated Rician channels. To the best of the author{'}s knowledge, only asymptotic expressions have been given in the literature for the \ac{BER} and the outage probability ($P_{{\rm out}}$) for the general case \cite{Mallik2011, Ma2005} and infinite series representations when the number of branches is limited to $P=2$ \cite{Bithas2007, Ilic2013}. In the following, expressions for both the \ac{BER} and $P_{\rm out}$ are provided using the new approach here presented.

\subsection{System model}

Consider a MRC system with $P$ branches at the receiver side. Then, the received signal can be written as
\begin{equation}
	\mathbf{r} = \mathbf{g} z + \mathbf{w}
\end{equation}
where $z$ is the complex transmitted symbol with $\mathbb{E}\left[\left|z\right|^2\right] = E_s$, $\mathbf{w}\in\mathbb{C}^{P\times 1}$ is the noise vector, and $\mathbf{g}\in\mathbb{C}^{P\times 1}$ is the normalized channel complex gain vector. The noise at each branch is assumed to be independent and identically distributed with zero-mean and variance $N_0$. Since the fading at each branch is assumed to be Rician distributed with $K_i$ factor for $i = 1,\dots,P$, $\mathbf{g}$ is a complex Gaussian vector such that $\mathbf{g}\sim\mathcal{CN}_P\left(\mathbf{\overline{g}},\mathbf{\Sigma}\right)$, with $\mathbf{\overline{g}}=\mathbb{E}[\mathbf{g}]$ and $\mathbf{\Sigma}$ the covariance matrix. The entries of both the mean vector and the covariance matrix can be expressed in terms of the Rician factors as
\begin{equation}
	\overline{g}_i = \sqrt{\frac{K_i}{K_i+1}}, \quad \Sigma_{i,j} = \sqrt{\frac{1}{K_iK_j}}R_{i,j}
\end{equation}
with $R_{i,j}$ for $i,j = 1,\dots,P$ the entries of the correlation matrix $\mathbf{R}$ of $\mathbf{g}$. Note that each element of $\mathbf{g}$ has unit power, i.e. $\mathbb{E}\left[\left|g_i\right|^2\right] = 1$, so the average \ac{SNR} at each branch is given by $\overline{\gamma} = E_s/N_0$. 
Considering perfect symbol synchronization and channel estimation, when the \ac{MRC} principle is applied to the received signal, this yields to a post-processing signal that can be expressed as
\begin{equation}
	r_{\rm out} = \sum_{k=1}^P r_k\frac{g_k^{\dagger}}{\mathbf{g}^{\dagger}\mathbf{g}} =  z + \sum_{k=1}^P \frac{g_k^{\dagger}w_k}{\mathbf{g}^{\dagger}\mathbf{g}}.
\end{equation}

Thus, the post-processing \ac{SNR} is given by
\begin{equation}
	\label{eq:SNR}
	\gamma = \overline{\gamma}\,\mathbf{g}^{\dagger}\mathbf{g}.
\end{equation}

Since \eqref{eq:SNR} is the non-central CGQF given in (\ref{eq:QF}) with $\mathbf{A}=\mathbf{I}$, the theoretical results derived in this paper can be used to analyze the performance analysis of such system. To that end, it is necessary to define an auxiliary variable $\gamma_m$ as in \eqref{eq:QFconf} such as
\begin{equation}
	\gamma_m = \overline{\gamma}\left(\mathbf{y} + \mathbf{D}_{\xi}\mathbf{\overline{h}}\right)^{\dagger}\mathbf{\Lambda}\left(\mathbf{y} + \mathbf{D}_{\xi}\mathbf{\overline{h}}\right),
	\label{eq:SNRconf}
\end{equation} 
where $\mathbf{y}\sim\mathcal{CN}_P\left(\mathbf{0}_{P\times 1},\mathbf{I}_{P}\right)$, $\mathbf{\overline{h}} = \mathbf{U}^{\dagger}\mathbf{C}^{-1}\mathbf{\overline{g}} $ with $\mathbf{U}$ and $\mathbf{\Lambda}$ being the unitary matrix and the diagonal matrix built with the eigenvalues of $\mathbf{C}^{\dagger}\mathbf{C}$, respectively, with $\mathbf{\Sigma} = \mathbf{CC}^{\dagger}$. When defining $\mathbf{D}_{\xi}$ as in Section \ref{subsec:ConfCGQF}, $\gamma_m$ is confluent to $\gamma$ by virtue of Proposition  \ref{pro:MGF}. Therefore, the performance analysis of the system will be based on the characterization of $\gamma_m$ when $m$ takes appropriate large values.

\subsection{Outage probability}

Defining $\gamma_{\rm th}$ as the minimum \ac{SNR} required for a reliable communication, the outage probability is given by \cite[eq. (6.46)]{Goldsmith05}
\begin{equation}
	 P_{\rm out}(\gamma_{\rm th}) = P(\gamma < \gamma_{\rm th}) \approx P(\gamma_m < \gamma_{\rm th}) =  \int_0^{\gamma_{\rm th}}f_{\gamma_m}(\gamma_m)\;d\gamma_m,
\end{equation}
which corresponds to the CDF of $\gamma_m$.  Moreover, since $\mathbf{\Lambda}$ is positive definite, then $\widetilde{\beta}_i > 0\;\forall\; i$, such that the outage probability is written as
\begin{equation}
	\label{eq:Pout}
	P_{\rm out}(\gamma_{\rm th}) \approx 1 + \sum_{i=1}^{n_{\beta}}\sum_{j=1}^{p_i}\omega_{i,j}e^{-\widetilde{\beta}_i \gamma_{th}/\overline{\gamma}}\left(\frac{\overline{\gamma}}{\gamma_{{\rm th}}}\right)^{-j+1}
\end{equation}
where $\omega_{i,j}$, $\widetilde{\beta}_k$, $n_{\beta}$ and $p_i$ are defined in Section \ref{sec:StatChar}.

\subsection{BER for M-QAM} 

Since the \ac{BER} is a continuous and bounded function, by virtue of Lemma \ref{lem:ExpConv} it is possible to approximate the \ac{BER} over the SNR variable $\gamma$ through the analysis of the confluent variable $\gamma_m$. Therefore, assuming a Gray coded constellation, the exact \ac{BER} expression conditioned to a certain $\gamma_m$ for arbitrary $M$-ary square QAM is given by \cite{Lopez2010}
\begin{equation}
	\label{eq:PbGammaM}
	P_b(\gamma_m) = L\sum_{i=1}^{L-1}\omega(i)\;Q\left((2i-1)\sqrt{\frac{3\gamma_m}{M-1}}\right) 
\end{equation}
where $\omega(i)$ are constants defined in \cite[eq. (6), (14) and (21)]{Lopez2010}, $L = \sqrt{M}$ and $Q(\cdot)$ is the Gaussian $Q$-function \citep[eq. (4.1)]{Alouini05}. In order to obtain the \ac{BER} for the system model described in the previous section over correlated Rician channels, \eqref{eq:PbGammaM} is averaged over the distribution of $\gamma_m$, such as
\begin{equation}
	\label{eq:PbInt}
	P_{b}\left(\overline{\gamma}\right) \approx \int_0^{\infty}P_b\left(\gamma_m\right)f_{\gamma_m}(\gamma_m)\;d\gamma_m.
\end{equation}

From \eqref{eq:PbGammaM} and \eqref{eq:PbInt}, and using the relation between the Gaussian $Q$-function and the error function ${\rm erf}(\cdot)$ given in \cite[eq. (8.250 1)]{Gradshteyn07}, the average \ac{BER} is calculated by applying \cite[eq. (3.381 4)]{Gradshteyn07} and \cite[eq. (4.3.8)]{Edward69}, obtaining
\begin{equation}
	\label{eq:BER}
	P_b \left(\overline{\gamma}\right) \approx \sum_{i=1}^{n_\beta}\sum_{j=1}^{p_i}\sum_{k=1}^{L-1}L\,\omega(k)\alpha_{i,j}\left[\frac{\Gamma(j)}{2\widetilde{\beta}_i^{j}} - \frac{\delta_k\Gamma\left(j+\frac{1}{2}\right)}{\widetilde{\beta}_i^{j+1/2}}\sqrt{\frac{\overline{\gamma}}{2\pi}}{}_2F_1\left(\frac{1}{2}, j+\frac{1}{2}; \frac{3}{2};\frac{-\delta_k^2 }{2\widetilde{\beta}_i}\overline{\gamma}\right)\right]
\end{equation}
where $\delta_k = (2k-1)\sqrt{3/(M-1)}$ and ${}_2F_1(\cdot)$ is the Gauss hypergeometric function \cite[eq. (15.1.1)]{Abra72}. 

\subsection{Numerical Results}

In the following, the influence of the channel parameters and the number of branches of the receiver in the outage probability and the \ac{BER} is assessed using \eqref{eq:Pout} and \eqref{eq:BER} and contrasted through Monte-Carlo simulations. Although the theoretical expressions in \eqref{eq:Pout} and \eqref{eq:BER} were derived using the confluent \ac{SNR} $\gamma_m$ in \eqref{eq:SNRconf}, the original variable $\gamma$ in \eqref{eq:SNR} is used in the simulations in order to validate the accuracy of the approximation. For the sake of simplicity, the vector containing the $P$ Rician $K$ factors is denoted as $\mathbf{k} = [K_1,\dots,K_P]$. Also, the correlation matrix $\mathbf{R}$ is assumed to be exponential, i.e. $\left(\mathbf{R}\right)_{i,j} = \rho^{\left|i-j\right|}$ with $\left|\rho\right| < 1$ \cite{Karagiannidis2003, Subadar2015, Loyka2001}. A thorough study has been performed by considering multiple combinations of $\rho$, $\mathbf{k}$ and the number of branches, $P$, which are varied over a large range of \ac{SNR}. While a detailed analysis of the results, depicted in Figs. \ref{fig:1}-\ref{fig:6}, is given below, it is important to notice that there is a perfect match between the analytical and the simulated values in all cases.  

Firstly, the impact of the correlation matrix and the Rician $K$ factors in the outage probability is studied both in the low-SNR and high-SNR regime. Since $P_{\rm out}$ exhibits complementary behaviors in both regimes, a different representation is employed in each case. Fig. \ref{fig:1} depicts the complementary outage probability ($1-P_{\rm out}$) when the SNR takes low values compared to the threshold, whereas Fig. \ref{fig:2} show the values of $P_{\rm out}$ in the high-SNR regime. In both cases the number of branches at the receiver is fixed to $P=2$. Note that the correlation between branches and the strength of the \ac{LoS} have opposite effects in the low and high SNR regimes. Hence, while in the latter a strong \ac{LoS} achieves a better performance than a weak one, in the low \ac{SNR} range a weak direct component seems to be beneficial. Similarly, while a high correlation factor gives better performance than a low one when $\overline{\gamma}$ is large, the opposite behavior is observed for low values of $\overline{\gamma}$. Altough similar conclusions are given in \cite{Bithas2007, Ilic2013} for $P_{\rm out}$ in the high \ac{SNR} regime, no attention have been paid in the literature to the behavior of the outage probability when the \ac{SNR} takes very low values compared to the threshold.

\begin{minipage}{\linewidth}
      \centering
      \begin{minipage}{0.45\linewidth}
          \begin{figure}[H]
              \includegraphics[width=\linewidth]{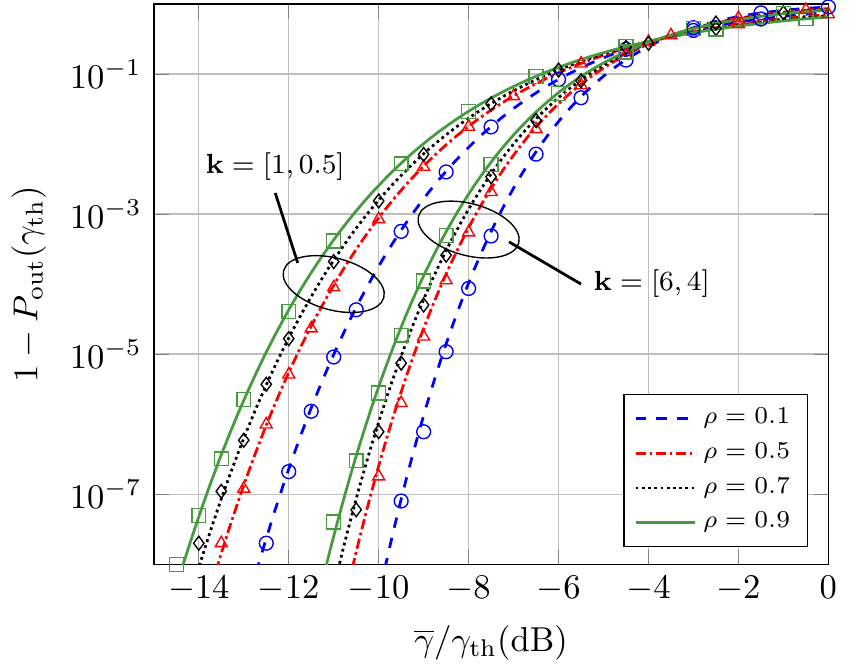}
              \caption{Complementary $P_{\rm out}$ vs. $\overline{\gamma}/\gamma_{\rm th}$ for $P = 2$, different values of $\rho$ and different values of $K$ at each path. Solid lines correspond to theoretical calculation with $m=40$ for $\mathbf{k}=[1,0.5]$ and $m=100$ for $\mathbf{k}=[6,4]$, while markers correspond to Monte Carlo simulations.}
              \label{fig:1}
          \end{figure}
      \end{minipage}
      \hspace{0.05\linewidth}
      \begin{minipage}{0.45\linewidth}
          \begin{figure}[H]
              \includegraphics[width=\linewidth]{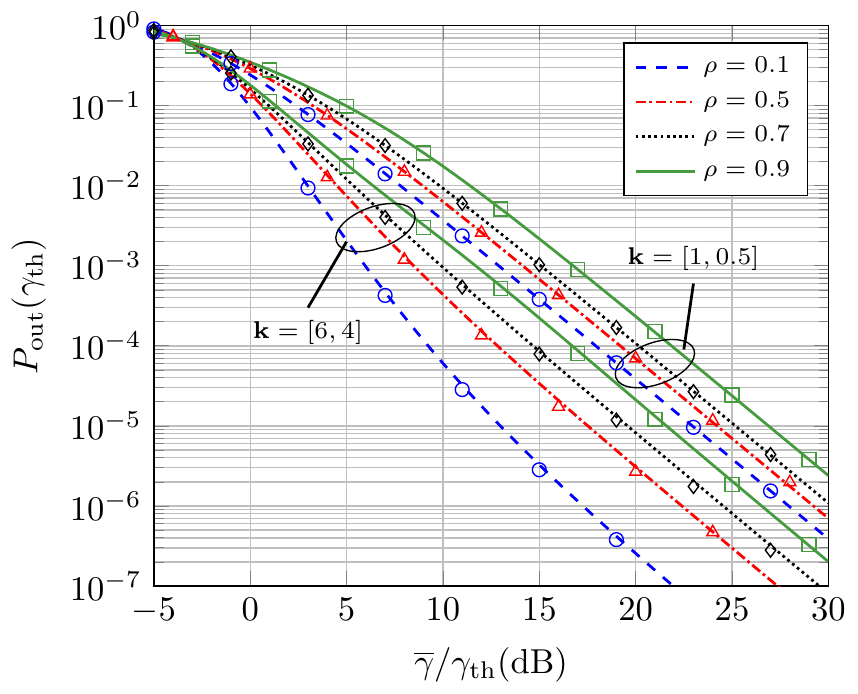}
              \caption{$P_{\rm out}$ vs. $\overline{\gamma}/\gamma_{\rm th}$ for $P = 2$, different values of $\rho$ and different values of $K$ at each path. Solid lines correspond to theoretical calculation with $m=50$ for $\mathbf{k}=[1,0.5]$ and $m=150$ for $\mathbf{k}=[6,4]$, while markers correspond to Monte Carlo simulations.}
              \label{fig:2}
          \end{figure}
      \end{minipage}
\end{minipage}

The analysis of the outage probability now focuses on the high-SNR regime. The number of branches is extended to $P=4$ to enrich the system. Fig. \ref{fig:3} assesses the influence of the correlation factor in strong and weak \ac{LoS} scenarios. It can be observed that the impact of  $\rho$ depends on the strength of the \ac{LoS}. Hence, increasing the correlation between branches implies a considerable degradation of the system performance when the \ac{LoS} is strong (large $K_i$ factors). In fact, when $\rho = 0.9$, the outage probability is asymptotically higher in a strong \ac{LoS} scenario than in a weak one. This effect is analyzed in more detail in Fig. \ref{fig:4}, where $P_{\rm out}$ is plotted for different vales of $\mathbf{k}$ when $\rho = 0.1$ (low correlation between branches) and $\rho = 0.9$ (branches highly correlated). As seen, the system behaves as expected when $\rho = 0.1$, since $P_{\rm out}$ decreases as the entries of $\mathbf{k}$ increases. However, when $\rho =0.9$, the system performance does not monotonically improves with the strength of the \ac{LoS}. Only when the distinct $K_i$ factors reach a certain value,  $P_{\rm out}$ decreases as the $K_i$ factors increase. The value of this turning point seems to depend on the number of branches and the correlation between them. This behavior was deeply analyzed in \cite{Wu2016}, where a \ac{MIMO-MRC} system is considered, providing expressions for this threshold value for the different $K_i$ factors. 

\begin{minipage}{\linewidth}
      \centering
      \begin{minipage}{0.45\linewidth}
          \begin{figure}[H]
              \includegraphics[width=\linewidth]{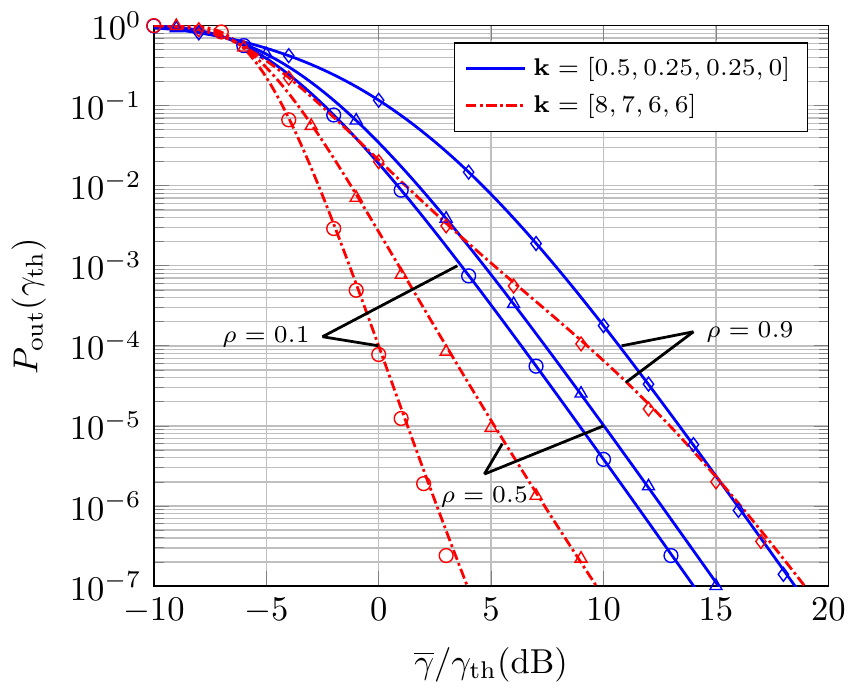}
              \caption{$P_{\rm out}$ vs. $\overline{\gamma}/\gamma_{\rm th}$ for $P = 4$, different values of $\rho$ and different values of $K$ at each path. Solid lines correspond to theoretical $P_{\rm out}$ while markers correspond to Monte Carlo simulations. For theoretical calculation, $m=40$ for $\mathbf{k} = [0.5,0.25,0.25,0]$ and $m=200$ for $\mathbf{k} = [8,7,6,6]$.}
              \label{fig:3}
          \end{figure}
      \end{minipage}
      \hspace{0.05\linewidth}
      \begin{minipage}{0.45\linewidth}
          \begin{figure}[H]
              \includegraphics[width=\linewidth]{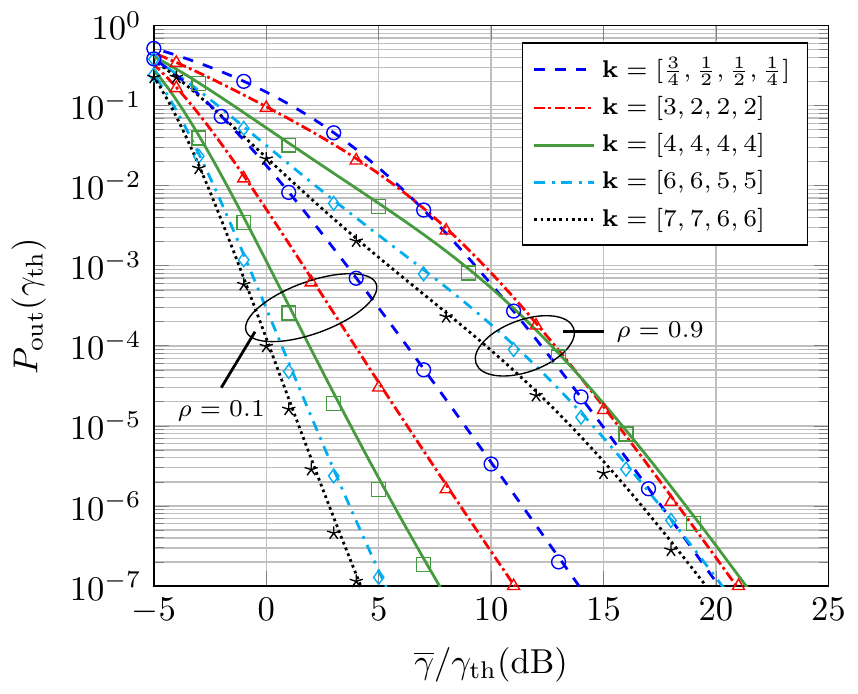}
              \caption{$P_{\rm out}$ vs. $\overline{\gamma}/\gamma_{\rm th}$ for $P = 4$ and different values of $\mathbf{k}$ with a strong correlation factor ($\rho=0.9$) and a weak correlation factor ($\rho=0.1$). Solid lines correspond to theoretical $P_{\rm out}$ while markers correspond to Monte Carlo simulations. For theoretical calculation, $m\in[40,200]$.}
              \label{fig:4}
          \end{figure}
      \end{minipage}
\end{minipage}

Regarding the \ac{BER}, the impact of the correlation factor and of the strength of the \ac{LoS} is firstly evaluated. Fig. \ref{fig:5} depicts the \ac{BER} for $16$-QAM with different values of $\rho$ in strong and weak \ac{LoS} scenarios. The influence of the modulation scheme is appraised in Fig. \ref{fig:6}, where the \ac{BER} for varios QAM constellations and different values of $\mathbf{k}$ are represented for a fixed correlation factor. It is interesting to observe that the \ac{BER} suffers the same relative degradation in all modulation schemes when the $K_i$ factors decrease. Similar conclusions were drawn in  \cite{Haghani2007,Haghani2008} when different reception techniques are applied over correlated Rician channels.

\begin{minipage}{\linewidth}
      \centering
      \begin{minipage}{0.45\linewidth}
          \begin{figure}[H]
              \includegraphics[width=\linewidth]{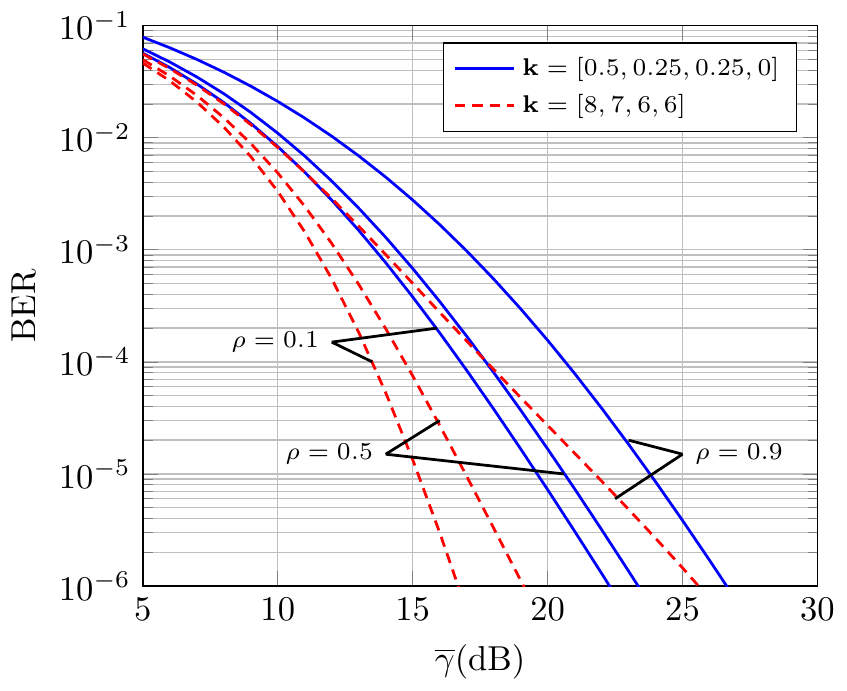}
              \caption{BER vs. $\overline{\gamma}$ for $16$-QAM, $P = 4$ and different values of $\rho$ and $\mathbf{k}$. Solid lines correspond to theoretical BER with $m=40$ for $\mathbf{k} = [0.5,0.25,0.25,0]$ and $m=150$ for $\mathbf{k} = [8,7,6,6]$, while markers correspond to Monte Carlo simulations.}
              \label{fig:5}
          \end{figure}
      \end{minipage}
      \hspace{0.05\linewidth}
      \begin{minipage}{0.45\linewidth}
          \begin{figure}[H]
              \includegraphics[width=\linewidth]{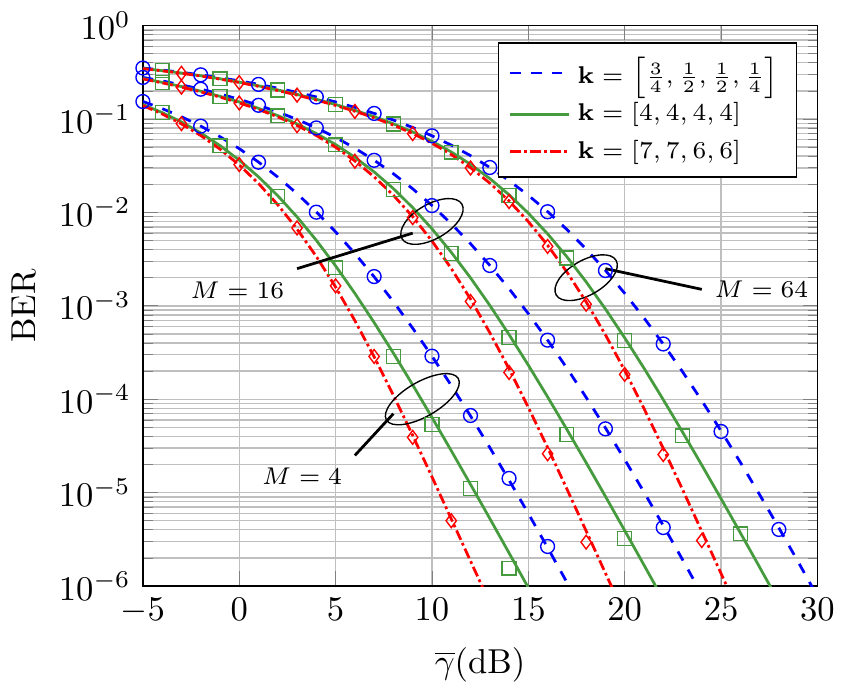}
              \caption{BER vs. $\overline{\gamma}$ for different modulation schemes and different values of $\mathbf{k}$ with $P=4$ and $\rho = 0.5$. Solid lines correspond to theoretical BER while markers correspond to Monte Carlo simulations. For theoretical calculation, $m\in[50,150]$.}
              \label{fig:6}
          \end{figure}
      \end{minipage}
\end{minipage}

Results displayed in Figs. \ref{fig:1}-\ref{fig:6} have been obtained with different values of the shape parameter $m$. As the latter increases, the \ac{MSE} between the confluent CGQF and the original one decreases, rendering a better accuracy in the approximation. However, the value of $m$ required to achieve a given \ac{MSE} depends on the characteristics of the CGQF. Fig. \ref{fig:7} shows the \ac{MSE} given in \eqref{eq:MSE}, normalized by $\Omega = \mathbb{E}\left[\gamma^2\right]$. The \ac{MSE} is always below $10^{-2}$ for the values of $m$ used in the theoretical calculations, which justifies the good match with the simulations. Note also that larger $K_i$ factors and higher correlation between branches require larger values of $m$ to reach a certain \ac{MSE}. Interestingly, the slope of the \ac{MSE} does not depend on the channel parameters, being the same for all the cases.  

\begin{minipage}{\linewidth}
      \centering
          \begin{figure}[H]
          		\centering
              \includegraphics[width=0.45\linewidth]{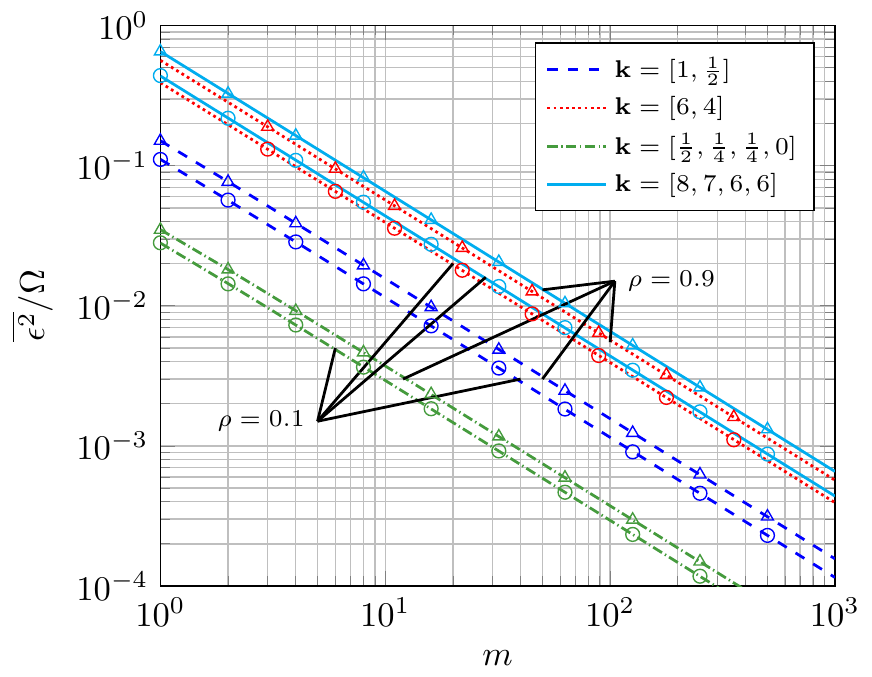}
              \caption{Normalized MSE vs. $m$ for different values of $P$, $\rho$ and $\mathbf{k}$. Solid lines correspond to theoretical MSE while markers correspond to Monte Carlo simulations.}
              \label{fig:7}
          \end{figure}
\end{minipage}

\section{Conclusions}
\label{sec:Conclusion}
This paper has presented a novel approach to the statistical characterization of indefinite non-central \ac{CGQFs}. Its key idea is to perturb the non-central vector of the CGQF with a random variable that depends on a shape parameter. The resulting auxiliary CGQF, which converges to the original one when this parameter tends to infinity, has simpler \ac{PDF} and \ac{CDF} expressions. In contrast to previous approaches available in the literature, results derived herein permits further insightful analyses, since the resulting probability functions are expressed in terms of elementary functions (exponential and powers) that can be used in subsequent calculations. Also, the \ac{MSE} between the auxiliary CGQF and the original one is given in closed-form, allowing the particularization of the auxiliary CGQF in order to make this error drop below a certain threshold. 

The usefulness of the proposed method has been exemplified by the analysis of \ac{MRC} systems over non-identically distributed Rician fading channels with arbitrary correlation, whose outage probability and \ac{BER} expressions are given and validated through Monte-Carlo simulations, showing a perfect match between the theoretical calculations and the simulations.

%

\appendices

\section{Proof of Proposition \ref{pro:MGF}}
\label{app:MGFproof}

Consider the confluent CGQF $Q_m$ defined in \eqref{eq:QFconf}. When conditioned on $\mathbf{D_{\xi}}$, its MGF is obtained from \eqref{eq:MGFCGQF} as
\begin{equation}
	\label{eq:condMGF}
	M_{Q_m\rvert_{\mathbf{D_{\xi}}}}(s) = \prod_{i=1}^n\frac{\exp\left({\dfrac{\xi_{m,i}^2\lambda_i\mu_is}{1-\lambda_is}}\right)}{1-\lambda_is},
\end{equation}

The unconditional \ac{MGF} is obtained by integrating over each variable $\xi_{m,i}^2$ as
\begin{equation}
	\label{eq:MGFintegral}
	M_{Q_m}(s) = \int_0^{\infty}\cdots\int_0^{\infty}{M_{Q_m\rvert_{\mathbf{D_{\xi}}}}(s)\,f_{\xi^2_{m,1},\dots \xi^2_{m,n}}(u_1,\dots,u_n)\,du_1\dots du_n}.
\end{equation}
with $f_{\xi_{m,1}^2,\dots,\xi_{m,n}^2}(\cdot)$ the joint probability density function of $\xi_{m,1}^2,\dots,\xi_{m,n}^2$. Since $\xi_{m,i}^2$ for $i=1,\dots,n$ are independent random variables, their joint density function can be calculated as
\begin{equation}
	f_{\xi^2_{m,1},\dots \xi^2_{m,n}}(u_1,\dots,u_n) = \prod_{i=1}^n{f_{\xi_{m,i}^2}(u_i)},
\end{equation}
where $f_{\xi_{m,i}^2}(u_i)$ for $i=1,\dots,n$ is the PDF of the Gamma distribution with shape parameter $m$ and scale parameter $1/m$, given in \eqref{eq:GammaPDF}. By substituting in \eqref{eq:MGFintegral}, the MGF of the confluent CGQF is rewritten as follows
\begin{equation}
	M_{Q_m}(s) = \prod_{i=1}^n{\frac{m^m}{\Gamma(m)(1-\lambda_is)}}\int_0^{\infty}\cdots\int_0^{\infty}{\prod_{k=1}^n{\exp\left(-u_k\left[m - \frac{\lambda_k\mu_ks}{1-\lambda_ks}\right]\right)}\,du_1\dotsc du_n}
\end{equation}
where, due to the independence between variables, each integral can be solved separately using \cite[eq. 3.381 4]{Gradshteyn07}, which yields to \eqref{eq:MGF} after some algebraic manipulations.

\section{Derivation of $A_{i,j}$ and $C_{i,j}$}
\label{app:Residues}

$A_{i,j}$ for $i = 1,\dots,n_\beta$ and $j=1,\dots,p_i$ arise as the partial expansion residues of the rational polynomial in \eqref{eq:MGFRsSimplf} after evaluating $M_{Q_m}(-s)$, whose general expression is given in terms of derivatives of the rational polynomial as \cite[eq. (A.36)]{Oppenheim96}
\begin{equation}
	A_{i,j} = \frac{1}{(p_i-j)!}\left.\frac{d^{p_i-j}}{ds^{p_i-j}}\left(\frac{\prod\limits^{n_\lambda}_{t=1}{\left(s+1/{\widetilde{\lambda}_t}\right)^{q_t}}}{\prod\limits^{n_\beta}_{\substack{l=1\\ l\neq i}}{\left(s+\widetilde{\beta}_l\right)^{p_l}}}\right)\right|_{s=-\widetilde{\beta}_i}.	\label{eq:AijDer} 
\end{equation}

By using the generalization of  Leibniz{'}s rule, the rational polynomial derivatives can be rewritten in term of the derivatives of the individual binomials as 
\begin{align}
	A_{i,j} =& \sum_{k_1+...+k_{N-1}=p_i-j}\left(\frac{1}{\prod\limits_{s=1}^{N}k_s}\prod_{t=1}^{n_\lambda}\left[\left(s+\widetilde{\lambda}^{-1}\right)^{q_t}\right]^{(k_t)} \prod_{r=1}^{i-1}\left[\left(s+\widetilde{\beta}_r\right)^{-p_r}\right]^{(k_{r+n})} \right. \notag \\
&\times\left.\left.\prod_{l=i+1}^{n_{\beta}}{\left[\left(s+\widetilde{\beta}_l\right)^{-p_l}\right]^{(k_{l+n-1})}}\vphantom{\frac{1}{\prod\limits_{s=1}^{N}k_s}}\right)\right|_{s=-\widetilde{\beta}_i}
	\label{eq:Aij_rule}
\end{align}
where the sum is over all possible combinations of $k_1,...,k_{N-1}$, with $N = n_\lambda + n_\beta$, that meet \mbox{$\sum\limits_{t=1}^{N-1}k_t=p_i-j$}. Thus, $A_{i,j}$ for $i = 1,\dots,n_\beta$ and $j=1,\dots,p_i$ are expressed as a finite sum of the $q$-th derivative of binomials with positive and negative exponents, which can be written in closed-form by using
\begin{equation}
	\label{eq:Ap4_1}
	\frac{d^q}{dx^q}(x+a)^{\nu} = 
	\begin{mycases}
	\frac{(-1)^q(-\nu)_q}{(x+a)^{-\nu+q}}, & \text{if} \;\; \nu < 0 \\
	\frac{\nu!}{(\nu-q)!}(x+a)^{\nu-q}, & \text{if} \;\; \nu > 0
	\end{mycases}.
\end{equation}
Then, the final expression for $A_{i,j}$ is given in \eqref{eq:Aij}, which has been obtained from \eqref{eq:Aij_rule} and \eqref{eq:Ap4_1} after evaluating at $s=-\widetilde{\beta}_i$. Note that, if $\nu > 0$, \eqref{eq:Ap4_1} is only valid for $q\leq\nu$ since the derivative is zero otherwise. Therefore, the restriction $k_u \leq q_u, u=1,\dots,n_\lambda$ is imposed in \eqref{eq:Aij}. 

Analogously, $C_{i,j}$ are the partial expansion residues of $R(s)/s$, so following the same steps as with $A_{i,j}$ one gets \eqref{eq:Cij}.

\section{Proof of Proposition \ref{pro:PDF}}
\label{app:PDFproof}

The PDF of $Q_m$ is obtained by performing an inverse Laplace transformation to the MGF such as
\begin{equation}
	f_{Q_m}(x) = \mathcal{L}^{-1}\left\{M_{Q_m}(-s)\right\}.
\end{equation}

Thus, evaluating \eqref{eq:MGFRsSimplf} at $-s$  and performing a partial fraction decomposition, one has
\begin{equation}
	M_{Q_m}(-s) = \prod_{k=1}^n{\left[\lambda_k\left(1+\frac{\mu_k}{m}\right)^m\right]^{-1}}\sum_{i=1}^{n_{\beta}}\sum_{j=1}^{p_i}{\frac{A_{i,j}}{\left(s+\widetilde{\beta}_i\right)^j}} \label{eq:MGF_PDFfrac}
\end{equation}
with $A_{i,j}$ the partial fraction decomposition residues given in \eqref{eq:Aij}, which are deduced in Appendix \ref{app:Residues}. The expression of the \ac{PDF} is easily derived from above equation just applying the Laplace transform pair \cite[p. 692]{Oppenheim96}
\begin{equation}
	\label{eq:LaplacePair}
	\mathcal{L}^{-1}\left\{\frac{1}{(s+\alpha)^\nu}\right\} = \begin{mycases}
		\frac{t^{\nu-1}}{(\nu-1)!}e^{-\alpha t}{\rm u}(t), & \text{if} \quad \alpha\geq 0 \\
		\frac{-t^{\nu-1}}{(\nu-1)!}e^{-\alpha t}{\rm u}(-t), & \text{if} \quad \alpha <0
	\end{mycases},
\end{equation}
which yields to \eqref{eq:PDF} after further algebraic manipulations.

\section{Proof of Proposition \ref{pro:CDF}}
\label{app:CDFproof}

The CDF of $Q_m$ is obtained from the MGF as 
\begin{equation}
	F_{Q_m}(t) = \mathcal{L}^{-1}\left\{\frac{1}{s}M_{Q_m}(-s)\right\}.
\end{equation}

Similarly as in Appendix \ref{app:PDFproof}, after performing a partial fraction expansion one has
\begin{equation}
	\label{eq:Ap3_1}
	\frac{1}{s}M_{Q_m}(-s) = \frac{1}{s}+\prod_{k=1}^n{\left[\lambda_k\left(1+\frac{\mu_k}{m}\right)^m\right]^{-1}}\sum_{i=1}^{n_{\beta}}\sum_{j=1}^{p_i}{\frac{C_{i,j}}{\left(s+\widetilde{\beta}_j\right)^j}}.
\end{equation}
where $C_{i,j}$ are the partial expansion residues given in \eqref{eq:Cij}, whose proof can be deduced from that of $A_{i,j}$ in Appendix \ref{app:Residues}. The final expression for the CDF in \eqref{eq:CDF} is straightforwardly obtained from \eqref{eq:Ap3_1} by applying the Laplace transform pair shown in \eqref{eq:LaplacePair}.

\section{Calculation of MSE}
\label{app:Error}

The \ac{MSE} between $Q_m$ and $Q$ is given by
\begin{equation}
	\label{eq:EpsilonDef}
	\overline{\epsilon^2} = \mathbb{E}\left[\left(Q_m-Q\right)^2\right] =  \mathbb{E}_{\mathbf{y},\mathbf{D}_{\xi}}\left[\left( \left(\mathbf{y}+\mathbf{D_{\xi}}\mathbf{\bar{h}}\right)^{\dagger}\mathbf{\Lambda}\left(\mathbf{y}+\mathbf{D_{\xi}}\mathbf{\bar{h}}\right) - \left(\mathbf{y}+\mathbf{\bar{h}}\right)^{\dagger}\mathbf{\Lambda}\left(\mathbf{y}+\mathbf{\bar{h}}\right)\right)^2\right].
\end{equation}

A simple way of performing the above expectation is considering first the \ac{MSE} conditioned to $\mathbf{D}_{\xi}$ (or equivalently, to $\xi_{m,i}$ for $i=1,\dots,n$). Expanding the square in \eqref{eq:EpsilonDef}, considering the expectation of a CGQF given in \cite[eq. (3.2b.2)]{Provost92} and taking into account the fact that the entries of $\mathbf{y}$ are mutually independent zero-mean complex Gaussian random variables whose real and imaginary parts are also independent and identically distributed, the conditioned \ac{MSE} is expressed as
\begin{align}
	\label{eq:MSEcondRes}
	\left.\overline{\epsilon^2}\right|_{\mathbf{D_{\xi}}} =&{\rm tr}\left(\mathbf{\Lambda}\left(\mathbf{D}_{\xi}-\mathbf{I}\right)\mathbf{\overline{h}\overline{h}}^{\dagger}\left(\mathbf{D}_{\xi}-\mathbf{I}\right)\mathbf{\Lambda}\right) + \mathbf{\overline{h}}^{\dagger}\left(\mathbf{D}_{\xi}-\mathbf{I}\right)\mathbf{\Lambda^2}\left(\mathbf{D}_{\xi}-\mathbf{I}\right)\mathbf{\overline{h}} \notag \\
	&+ \mathbf{\overline{h}}^{\dagger}\left(\mathbf{D}_{\xi}^2-\mathbf{I}\right)\mathbf{\Lambda}\mathbf{\overline{h}}\mathbf{\overline{h}}^{\dagger}\left(\mathbf{D}_{\xi}^2-\mathbf{I}\right)\mathbf{\Lambda}\mathbf{\overline{h}}.
\end{align}

The unconditional \ac{MSE} is obtained by averaging \eqref{eq:MSEcondRes} as
\begin{equation}
	\label{eq:ErrCal}
	\overline{\epsilon^2} = \mathbb{E}_{\mathbf{D}_\xi}\left[\left.\epsilon^2\right|_{\mathbf{D_{\xi}}}\right].
\end{equation}
Considering that both $\mathbf{D}_{\xi}$ and $\mathbf{\Lambda}$ are diagonal matrices and $\mathbb{E}\left[\mathbf{D}_{\xi}^2\right] = \mathbf{I}_n$,
\begin{equation}
	\label{eq:Epsilon_last}
	\overline{\epsilon^2} = 4\left(1-\overline{\xi}\right)\sum_{i=1}^n\lambda_i^2\left|h_i\right|^2 + {\rm tr}\left(\mathbf{\Lambda\mathbf{\overline{hh}}^{\dagger}\mathbf{\Sigma}}\right)
\end{equation}
where $\overline{\xi} = \mathbb{E}\left[\xi_{m,i}\right]$ for $i=1,\dots ,n$ is the expectation of a Nakagami-$m$ random variable which is given by
\begin{equation}
	\overline{\xi} = \frac{\Gamma(m+1/2)}{m^{1/2}\Gamma(m)}
\end{equation}
and $\mathbf{\Sigma}$ is the covariance matrix of $\mathbf{x} = \mathbf{\mathbf{D}_{\xi}^2\overline{h}}$.  Due to the statistical independence of the elements of $\mathbf{x}$, $x_i$ for $i = 1,\dots,n$, $\mathbf{\Sigma}$ is a diagonal matrix whose entries are the variances of the entries of $\mathbf{x}$. Since $\xi_{m,i}^2\sim\Gamma\left(m,1/m\right)\; \forall \;i$, then ${\rm Var}[x_i] = \left|h_i\right|^2/m$, where $h_i$ for $i=1,\dots,n$ are the entries of $\mathbf{\overline{h}}$. The final expression for the \ac{MSE} in \eqref{eq:MSE} is obtained by substituting the value of $\mathbf{\Sigma}$ in \eqref{eq:Epsilon_last} and performing some algebraic manipulations.
  
\vspace{10mm}
\bibliographystyle{IEEEtran}
\bibliography{Bibliografia}

\end{document}